# A coherence interpretation of nonlocal realism in the delayed-choice quantum eraser

Byoung S. Ham

School of Electrical Engineering and Computer Science, Gwangju Institute of Science and Technology, 123 Chumdangwagi-ro, Buk-gu, Gwangju 61005, S. Korea

(Submitted on August 11, 2023; bham@gist.ac.kr)

**Abstract**
The delayed-choice thought experiment proposed by Wheeler has been demonstrated over the last several decades for the wave-particle duality in quantum mechanics. The delayed-choice quantum eraser proposed by Scully and Drühl has also been intensively studied for the violation of the cause-effect relation of classical physics. Here, a coherence analysis is presented for the observed quantum eraser to understand the fundamental physics of nonlocal realism. As a result, coherence solutions of the observed quantum eraser are analytically derived from coincidence detection-caused selective measurements, where the resulting product-basis superposition becomes the origin of the otherwise quantum mystery of the nonlocal fringes in a joint parameter relation. For this, a fixed sum-phase relation between paired photons is a prerequisite, which cannot be explained by conventional particle nature-based quantum mechanics.

## 1 Introduction

The delayed-choice thought experiment proposed by Wheeler in 1978 is for the fundamental quantum mechanics of the wave-particle duality of a single photon in an interferometric system [1]. The delayed-choice quantum eraser proposed by Scully and Drühl in 1982 is for the violation of the cause-effect relation whose delayed choice satisfies the space-like separation [2]. The first experimental proof of the quantum eraser was conducted in 2000 [3] in a double-slit system of a type-II spontaneous parametric down-conversion (SPDC) medium [4,5]. The purpose of the delayed-choice thought experiment is to verify the post-determination of a single photon's nature, violating the local realism of classical physics. Various kinds of photon natures and measurement techniques have been used to understand quantum mechanics [6-10]. For the quantum eraser, the violation of the cause-effect relation has been tested in a delayed-choice scheme for various quantum perspectives [11-15]. Both delayed-choice and quantum erasers are for the wave-particle duality using a single photon or paired photons. In the paired photon regime, entangled photons have also been applied for the quantum eraser to study the weirdness of nonlocal quantum correlation [3,7,10,12].

In this paper, a coherence analysis is presented for the delayed-choice quantum eraser observed in a nonlocal regime using SPDC-generated entangled photon pairs [3]. Here, the 'coherence' represents the wave nature of a photon, which is equivalent to a harmonic oscillator defined by Maxwell's wave equation. Thus, the coherent analysis is fully equivalent to conventional coherence optics, where the effective coherence time is determined by the inverse of the spectral bandwidth of an ensemble of SPDC-generated photons. In this photon ensemble whose spectral bandwidth is Gaussian distributed according to the second-order nonlinear optics, each individual photon's coherence time should be much longer than the effective one due simply to the inhomogeneous broadening of the Gaussian curve. The coherently obtained solutions of the observed nonlocal fringes for ref. [3] satisfy the joint-phase relation of the independent local parameters. This is contradictory to our common understanding of nonlocal correlation which is known to be impossible by any classical means. For the coherence approach, however, a measurement modification is necessary for the nonlocal fringes. This selective measurement results in the second-order amplitude superposition between space-like separated photons comparable to the first-order amplitude superposition of a single photon [16].

## 2 Coherence interpretation

Figure 1 shows a schematic of the quantum eraser based on SPDC-generated photon pairs [3], whose theoretical model was proposed by Scully and Drühl in 1982 [2]. The paired photons are from a type-II SPDC nonlinear crystal BBO. The observed quantum correlation is between orthogonally polarized photons in a space-like separation regime [4]. In the type-II SPDC process, a phase-matching condition determines the phase relationship among the pump, signal, and idler photons according to the energy and momentum conservation laws [5]. A fixed sum-phase correlation between paired photons is the bedrock of the present coherence approach [17,18].

In Fig. 1(a), two open slits made on a BBO crystal are used to mimic conventional double slits, where SPDC-generated signal ($f_s$) and idler ($f_i$) photons are simultaneously excited from either slit by a pump photon whose frequency is $2f_0$. For this, the pump beam's diameter is adjusted to cover both slits [3, 19]. The resulting frequencies of the signal and idler photons are oppositely detuned across the half pump frequency $f_0$ by the energy conservation law, as shown in Fig. 1(b). Using a much narrower linewidth of a pump laser than $\Delta$ of the entangled photons, the following energy relation is satisfied: $2f_0 = f_s + f_i$. The bandwidth $\Delta$ is caused by various non-ideal phase matching conditions in the birefringent medium of BBO, resulting in a Gaussian-distributed inhomogeneous broadening of Fig. 1(b). Thus, each photon's coherence time in Fig. 1(b) should be much longer than the effective coherence time $\Delta^{-1}$ [17].

By the momentum conservation law, the angle between the signal (e1) and idler (e2) photons' propagation directions is depicted in Fig. 1(a) (see also the circled cross-sections of entangled photons in the Inset) [19]. For the quantum eraser experiments [3], the signal and idler photons are spatially and spectrally separated by a Glen Thomson prism located right after BBO (not shown), where e1 (e2) also stands for the vertical (horizontal) polarization basis of the signal (idler) photons. Here, the photon pair used for the quantum eraser [3] does not satisfy the entanglement relation $|En\rangle = (|s\rangle_1|id\rangle_2 + |id\rangle_1|s\rangle_2)/2$, where $s$ and $id$ stand for the signal and idler photons, respectively.

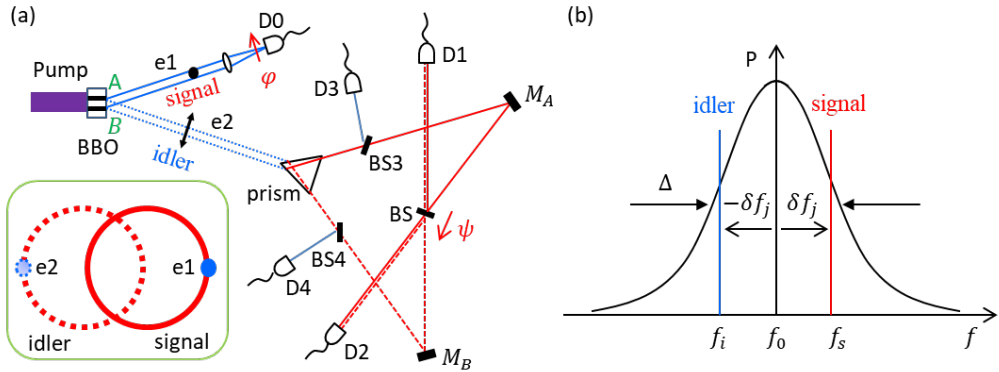

**Fig. 1. Schematic of the delayed-choice quantum eraser.** (a) Experimental setup. Inset: an example of phase-matched photon pair. (b) Frequency distribution of the $\chi^{(2)}$-generated photon pair. A/B: double slits, BS: balanced beam splitter, D: single photon detector, M: mirror. Inset: a cross-section of the $\chi^{(2)}$-generated photon pair and collection points of an entangled photon pair.

Satisfying the type-II SPDC process, all pairs of signal ($s$) and idler ($id$) photons in Fig. 1(a) have the same orthogonal polarization relation: $|s\rangle = |H\rangle$; $|id\rangle = |V\rangle$. By randomness of a pump photon falling in either slit on BBO, the signal (idler) photon in the e1 (e2) direction can be represented by the superposition of probability amplitudes from both slits A and B. Thus, the signal photon's amplitude $E_{10}$ along the e1 direction from the BBO crystal can be represented by $\mathbf{E}_{10}^j = \frac{E_0}{\sqrt{2}} e^{i\zeta_s}\big(\hat{V}_{Aj}^s + e^{i\eta_j}\hat{V}_{Bj}^s\big)$, where the relative phase $\eta_j$ between them is arbitrary due to independent slit-based spontaneous emission of nonlinear optics for different $f_s^j$ at different time. This understanding of nonlinear optics is quite important to proceed the present coherence analysis. Here, the ket (bra) notation in quantum mechanics is replaced by the electric field (conjugate) in coherence optics. $\zeta_s$ is a random global phase between entangled-photon pairs. Needless to say, the superposition relation of a signal photon for the detector D0 results in an incoherence feature for a mean intensity due to the random $\eta_j$.

For the type-II SPDC-generated idler photons along the e2 direction in Fig. 1(a), a prism is used to spatially distinguish the photon sources (slits A and B). As a result, the upper (lower) photon toward the mirror $M_A$ ($M_B$) is from the slit B (A). This spatial separation of idler photons is for the test of the quantum eraser along with signal photons for the two-photon intensity correlation. Here, the prism and BS for an idler photon configure an interferometer. Like the signal photons along the e1 direction, the idler photons from the double slits can be represented in the same way: $\mathbf{E}_{2k}^j = \frac{E_0}{\sqrt{2}} e^{i\zeta_{id}} (\hat{H}_{Aj}^{id} + e^{i\eta_j} \hat{H}_{Bj}^{id})$; k=1~4. Here, the random phase $\eta_j$ should be the same as the signal photon due to the simultaneous generation of entangled photons in SPDC. By the choice of BS3 (BS4), the intercepted idler photon is represented by $\hat{H}_{Bj}^{id}$ ($\hat{H}_{Aj}^{id}$) term in $\mathbf{E}_{2k}^j$, resulting in no chance of interference fringe due to no quantum superposition. This is the general understanding of the particle-nature concept for which-way information in quantum mechanics [3]. Combining idler photon's probability amplitudes on BS, however, satisfies self-interference of $\mathbf{E}_{2k}^j$. In this case, no ψ-dependent fringe is measured by D1 or D2 due to the random $\eta_j$. In a short summary, the local measurements by all detectors in Fig. 1(a) show uniform intensities due to either random phases or no interference.

Regarding the two-photon intensity correlation between D0 and D1, the observed nonlocal intensity fringe in ref. [3] is for orthogonally polarized photon pairs by definition of the type-II SPDC process. By the phase matching condition of $\chi^{(2)}$ nonlinear optics [5], the sum phase of each entangled photon pair should be dependent upon the pump photon's phase [13]. Because the pump photon's phase is random for different SPDC-generated photon pairs, the sum phase of each photon pair should be random, too. However, this does not mean that the relative phase between the signal and idler photons is also random. Instead, a fixed relative phase between entangled photons in each pair has already been theoretically analyzed for π/2 [18] in both Franson-type nonlocal correlation [17,20] and the Hong-Ou-Mandel effect [18,21]. Such a fixed phase relationship between the signal and idler photons has also been experimentally demonstrated for local fringes [12,22]. This fixed phase relation between entangled photons is the bedrock of the present coherent analysis for the observed nonlocal fringes in Ref. [3].

The break-up condition of the coherence between entangled photons to satisfy the space-like separation is not for individuals but for an ensemble of photon pairs with respect to the effective coherence $\Delta^{-1}$ [3]. This fact is quite important to apply the coherence approach to the cause-effect relation of the quantum eraser because the coherence condition is inherently satisfied between paired photons (see Discussion). As discussed in the Discussion, however, information should relate to the ensemble. In that information perspective, the space-like separation defined by $c\Delta^{-1}$ in ref. [3] makes sense for the violation of the cause-effect relation. Unlike uniform local intensities, the random phase $\eta_j$ does not play any role in the nonlocal intensity measurements due to the coincidence detection, where slits A and B must be separated in a time domain (see Analysis). The pump phase-dependent global phase between SPDC pairs has no effect, either, according to the Born rule of probability amplitude for individual and independent measurements [23].

## 3 Analysis

Using the superposition relation between two probability amplitudes of a j$^{th}$ photon generated from either slit A or B in Fig. 1(a), individual photon's amplitudes on detectors D0~D4 can be represented as:

$$\mathbf{E}_{10}^j = \frac{E_0}{\sqrt{2}} e^{i\zeta_s} (\hat{V}_{Aj}^s + e^{i(\eta_j + \varphi_j)} \hat{V}_{Bj}^s), \tag{1}$$

$$\mathbf{E}_{21}^j = \frac{E_0}{\sqrt{2}} e^{i\zeta_{id}} (\hat{H}_{Aj}^{id} + i e^{i(\eta_j - \psi_j)} \hat{H}_{Bj}^{id}), \tag{2}$$

$$\mathbf{E}_{22}^j = \frac{E_0}{\sqrt{2}} e^{i\zeta_{id}} (i\hat{H}_{Aj}^{id} + e^{i(\eta_j - \psi_j)} \hat{H}_{Bj}^{id}), \tag{3}$$

$$\mathbf{E}_{23}^j = \frac{E_0}{\sqrt{2}} e^{i\theta_{id}} \hat{H}_{Bj}^{id}, \tag{4}$$

$$\mathbf{E}_{24}^j = \frac{E_0}{\sqrt{2}} e^{i\theta_{id}} \hat{H}_{Aj}^{id}, \tag{5}$$

where $\hat{H}_{Zj}^{id} = \hat{h}e^{-i(\delta f_j t_{id})}$ and $\hat{V}_{Zj}^{s} = \hat{v}e^{i(\delta f_j t_s)}$ are vector notations in horizontal and vertical polarizations of the photon, respectively, from the source Z ∈ {A, B}Z. $E_0$ is the amplitude of the single photon. Here, the subscript *xy* in $\mathbf{E}_{xy}$ indicates the photon's propagation direction (x) and a target detector (y). The random phase $\eta_j$ is applied to the slit B-generated photon. $\zeta_s$ ($\zeta_{id}$) is an arbitrary global phase of each photon pair along the e1 (e2) direction. Likewise, $\theta_{id}$ is an arbitrary global phase of an idler photon toward D3 or D4. Due to the symmetric detuning between the signal and idler photons in Fig. 1(b), there is a $\pm$ sign change, resulting in $e^{\pm i(\delta f_j \tau)}$, respectively (not shown). As in double-slit experiments, the transversal movement-caused phase shift $\varphi$ should be replaced by $\varphi_j$ due to different $\delta f_j \tau$ for the *j*$^{th}$ photon from the B slit only in $\mathbf{E}_{10}^{j}$. Likewise, the BS scan-induced phase shift $\psi$ is replaced by $\psi_j$ for the *j*$^{th}$ photon from the B slit only in $\mathbf{E}_{21}^{j}$ and $\mathbf{E}_{22}^{j}$, reversing the $\pm$ sign in $e^{\pm i(\delta f_j \tau)}$ of $\mathbf{E}_{10}^{j}$. Thus, the following local intensities are coherently obtained:

$$\langle I_j \rangle_{10} = \langle \mathbf{E}_{10}^{j} (\mathbf{E}_{10}^{j})^* \rangle$$
$$= \frac{I_0}{2} \langle (\hat{V}_{Aj}^{s} + e^{i(\eta_j + \varphi_j)} \hat{V}_{Bj}^{s})(\hat{V}_{Aj}^{s} + e^{i(\eta_j + \varphi_j)} \hat{V}_{Bj}^{s})^* \rangle$$
$$= I_0 \langle 1 + \cos(\eta_j + \varphi_j) \rangle, \qquad (6)$$

$$\langle I_j \rangle_{21} = \langle \mathbf{E}_{21}^{j} (\mathbf{E}_{21}^{j})^* \rangle$$
$$= \frac{I_0}{2} \langle (\hat{H}_{Aj}^{id} + i e^{i(\eta_j - \psi_j)} \hat{H}_{Bj}^{id})(\hat{H}_{Aj}^{id} - i e^{i(\eta_j - \psi_j)} \hat{H}_{Bj}^{id})^* \rangle$$
$$= I_0 \langle 1 - \sin(\eta_j - \psi_j) \rangle, \qquad (7)$$

where $I_0 = E_0 E_0^*$. Likewise, $\langle I_j \rangle_{22} = I_0 \langle 1 + \sin(\eta_j - \psi_j) \rangle$. Here, $\langle X \rangle$ stands for the mean value of X measurements. Due to the random phase $\eta_j$ between photons from slits A and B, however, both cosine and sine terms become zero on average, regardless of $\varphi_j$ or $\psi_j$: $\langle I_j \rangle_{10} = \langle I_j \rangle_{21} = \langle I_j \rangle_{22} = I_0$. Thus, the observed uniform intensities in ref. [3] are analytically confirmed by the present coherence approach. This is the first validity of the present coherence analysis for the locally measured uniform intensities.

For the two-photon intensity correlation $R_{01}(\tau)$ between detectors D0 and D1, the concept of the second-order amplitude superposition is used for the product-basis selective measurements:

$$\langle R_j(0) \rangle_{01} = \langle \mathbf{E}_{10}^{j} \mathbf{E}_{21}^{j} (cc) \rangle$$
$$= \frac{I_0^2}{4} \left( \hat{V}_{Aj}^{s} + e^{i(\eta_j + \varphi_j)} \hat{V}_{Bj}^{s} \right) \left( \hat{H}_{Aj}^{id} + i e^{i(\eta_j - \psi_j)} \hat{H}_{Bj}^{id} \right) (cc)$$
$$= \frac{I_0^2}{4} \left( i e^{i(\eta_j - \psi_j)} \hat{V}_{Aj}^{s} \hat{H}_{Bj}^{id} + e^{i(\eta_j + \varphi_j)} \hat{V}_{Bj}^{s} \hat{H}_{Aj}^{id} \right) \left( i e^{i(\eta_j - \psi_j)} \hat{V}_{Aj}^{s} \hat{H}_{Bj}^{id} + e^{i(\eta_j + \varphi_j)} \hat{V}_{Bj}^{s} \hat{H}_{Aj}^{id} \right)^*$$
$$= \frac{I_0^2}{4} \left( (\hat{V}_{Aj}^{s} \hat{H}_{Bj}^{id})(\hat{V}_{Aj}^{s} \hat{H}_{Bj}^{id})^* + (\hat{V}_{Bj}^{s} \hat{H}_{Aj}^{id})(\hat{V}_{Bj}^{s} \hat{H}_{Aj}^{id})^* + i(\hat{V}_{Aj}^{s} \hat{H}_{Bj}^{id})(\hat{V}_{Bj}^{s} \hat{H}_{Aj}^{id})^* \left( e^{-i(\varphi_j + \psi_j)} - e^{i(\varphi_j + \psi_j)} \right) \right)$$
$$= \frac{I_0^2}{2} (1 - \sin(\varphi + \psi)), \qquad (8)$$

where $\tau = t_s - t_{id}$ and cc is a complex conjugate. Here, the detuning-independent relationship $\varphi_j + \psi_j = \varphi + \psi$ is due to the symmetric detuning between the signal and idler photons, as shown in Fig. 1(b). The coincidence detection allows only orthogonal product-basis terms by definition of the type-II SPDC process. This inherent property of type II SPDC-generated photons is the critical difference from the coherence optics whose intensity products allow all four tensor product bases. In other words, a measurement modification at the cost of 50% event loss for the same polarization product bases is a prerequisite in the viewpoint of coherence optics. This is the concept of the inevitable selective measurements in the present coherence approach [17]. Likewise, the coincidence detection-caused selective measurement process in Eq. (8) is the quintessence of the nonlocal correlation fringe, resulting in the joint-phase relation of local parameters $\varphi$ and $\psi$. This joint-phase relation is a unique quantum feature that cannot be obtained in classical physics [24]. Including the joint phase relation, understanding of the bandwidth (detuning)-independent nonlocal fringe is great benefits of the coherence approach.

Likewise, $\langle R_j(0) \rangle_{02} = \langle \mathbf{E}_{10}^{j} \mathbf{E}_{22}^{j} (cc) \rangle = I_0^2 (1 + \sin(\varphi + \psi))/2$ is also coherently obtained using Eqs. (1) and (3), as observed [3]. The opposite nonlocal fringes between $\langle R_{01}(\tau) \rangle$ and $\langle R_{02}(\tau) \rangle$ are the direct consequence of the

π/2 phase shift between the reflected and transmitted photons by the BS [25]. Thus, the nonlocal intensity-product fringes observed for the quantum eraser [3] are perfectly verified by the coherence analysis, where the derived joint-phase relation of the nonlocal fringes is an additional benefit of the complete coherence solution. This is the second validity of the coherence analysis for the nonlocal fringe of the quantum eraser observed in ref. [3].

Unlike coherently derived uniform intensities in Eqs. (6) and (7), the random phase $\eta_j$ between slits A and B in each measured photon pair does not affect the nonlocal intensity fringe in Eq. (8). This is because of the unique feature of the second-order nonlinear optics for the coincidence measurements. In other words, the coincidently measured nonlocal intensity is either from the slit A or B. Thus, the coherence solutions of the nonlocal intensity correlations are both bandwidth ($\Delta$) and $\eta_j$ independent. The mysterious quantum phenomenon of the nonlocal fringe in the quantum eraser [3] is now clearly and completely understood for the polarization product-basis superposition created by the polarization-basis selective measurements between controlled (idler) and uncontrolled (signal) photons via coincidence detection. The cost to pay for this nonlocal fringe in coherence optics of Eq. (8) is the 50% measurement-event loss. The orthogonal polarization product-basis superposition between paired photons indicates a new concept of the second-order amplitude superposition, which is an extension of the first-order amplitude superposition in a single photon's self-interference [16,26].

Now, we analyze the coincidence detection $\langle R_{03} \rangle$ between detectors D0 and D3:

$$\langle R_j(0) \rangle_{03} = \frac{I_0^2}{4} \langle \mathbf{E}_{10}^j \mathbf{E}_{23}^j (cc) \rangle$$
$$= \frac{I_0^2}{4} \left( \hat{V}_{Aj}^s + e^{i(\eta_j + \varphi_j)} \hat{V}_{Bj}^s \right) \left( \hat{H}_{Bj}^{id} \right) \left( \hat{V}_{Aj}^s + e^{i(\eta_j + \varphi_j)} \hat{V}_{Bj}^s \right)^* \left( \hat{H}_{Bj}^{id} \right)^*$$
$$= \frac{I_0^2}{4}. \tag{9}$$

Allowing photons from the slit B only for D3 by BS3, the uniform intensity in Eq. (9) is due to no superposition in coherence optics, which is equivalent to known which-way information in the particle nature of quantum mechanics. The coincidence detection in Eq. (9) also accompanies the 50 % event loss in normal coherence analysis for the selective choice of product bases rooted in coincidence measurements. Likewise, $\langle R_{04}(0) \rangle = I_0^2/4$ is also obtained between detectors D0 and D4 for the same reason. These uniform intensities in $\langle R_{03}(0) \rangle$ and $\langle R_{04}(0) \rangle$ are equivalent to the concept of distinguishable (which-way information) photon characteristics in the conventional particle-nature interpretations. Thus, the two-photon correlations derived in Eqs. (8) and (9) represent the distinguishable and indistinguishable photon characteristics of the quantum eraser by the "delayed" choice of BS and BS3, respectively. Due to the same coherence condition given to Eq. (8), the distinguishable photon characteristics in Eq. (9) are also confirmed by the same coherence analysis. Thus, the phase coherence in both distinguishable and indistinguishable photons corresponding to the wave-particle duality must be the bedrock of the nonlocal quantum feature. This coherent phase understanding of the quantum eraser is unique and distinctive with respect to conventional understandings of quantum mechanics.

## 4 Discussion

Regarding the space-like separation between the measured entangled photons by D0 and D1 as well as D0 and D2 for the delayed-choice quantum eraser [3], the causality violation is determined with respect to the speed of light. For the SPDC spectrum in Fig. 1(b), individual coherence time $\tau_j$ of each entangled photon pair should be much longer than the effective coherence time $\tau_0$ ($= \Delta^{-1}$) of the ensemble, i.e., $\tau_j \gg \tau_0$ by definition of Gaussian (inhomogeneous) distribution of SPDC-generated photon pairs. This coherence condition should redefine the space-like separation, which has never been considered yet carefully in the quantum information community [1-3,11]. Thus, such a coherence condition between local detectors by D0 and D1 in Eq. (8) seems to be contradictory to the nonlocal realism determined by the space-like separation between them [27,28]. This may be argued for further discussions of nonlocality, realism, and violation of causality. Like the long-term debate of superluminal information transfer [29,30], however, the information cannot be represented by a single phase of a monochromatic light but by a group (ensemble) phase by many waves. This is also the way to represent a quantum particle by a Fourier transform of many waves in a quantum superposition. Thus, the cause-effect relation in the present analysis of the coherence approach-based quantum eraser should be relieved by this information concept in terms of the ensemble coherence determined by the bandwidth Δ.

## 5  Conclusions

A delayed-choice quantum eraser observed in a double-slit system of a nonlinear crystal [3] was coherently analyzed for both jointly measured nonlocal intensity fringes and locally measured uniform intensities. As a result, the origin of the observed nonlocal fringe was found in the second-order amplitude superposition resulting from coincidence detection-caused selective measurements for polarization-product bases. Using coherence analysis, the resulting product-basis superposition between paired photons was shown to be the inherent random-phase independent nonlocal fringes. The coherently induced second-order amplitude superposition was new and even mysterious like the first-order amplitude superposition of a single photon resulted from the inherent polarization correlation in the SPDC process, otherwise coherently manipulated for a selective measurement process in the present coherence approach at the cost of 50% measurement event loss. Unlike the general understanding of the quantum eraser, the coincidently measured paired photons had to keep their coherence even for a delayed scheme to satisfy the product-basis superposition relation. For the locally measured uniform intensities, the coherence approach unveiled that a random phase between superposed photons from different slits of the nonlinear crystal was the origin. The derived coherence solution of the observed nonlocal fringes seemed to be contradictory to the violation of local realism because the 'phase coherence' between measured photons had to be preserved for the coherence regime. Considering the mean value of an ensemble of entangled photon pairs, however, the average of nonlocal correlation was meaningful for information determined by the effective coherence, as debated in the superluminal light. In that sense, the given condition of the space-like separation by the effective coherence might satisfy the violation of local realism even in the present coherence interpretation.

**Acknowledgments**
BSH thanks Y.-H. Kim at POSTECH for helpful discussions of ref. [3].

**Funding**
This work was supported by the MSIT (Ministry of Science and ICT), Korea, under the ITRC (Information Technology Research Center) support program (IITP-2023-2021-0-01810) supervised by the IITP (Institute for Information & Communications Technology Planning & Evaluation). BSH also acknowledges that this work was supported by GIST via GRI 2023.

**Competing interests**
The author declares no competing interests.